\documentclass[sigconf]{acmart}

\AtBeginDocument{%
	\providecommand\BibTeX{{%
			\normalfont B\kern-0.5em{\scshape i\kern-0.25em b}\kern-0.8em\TeX}}}

\usepackage{pdfpages}
\usepackage{libertine}
\usepackage{booktabs}
\AtBeginDocument{%
  \providecommand\BibTeX{{%
    \normalfont B\kern-0.5em{\scshape i\kern-0.25em b}\kern-0.8em\TeX}}}

\setcopyright{acmcopyright}

\acmSubmissionID{1222}

\begin{document}

\copyrightyear{2019}
\acmYear{2019}
\acmConference[RecSys '19]{Thirteenth ACM Conference on Recommender Systems}{September 16--20, 2019}{Copenhagen, Denmark}
\acmBooktitle{Thirteenth ACM Conference on Recommender Systems (RecSys '19), September 16--20, 2019, Copenhagen, Denmark}
\acmPrice{15.00}
\acmDOI{10.1145/3298689.3347067}
\acmISBN{978-1-4503-6243-6/19/09}

\title[On Gossip-based Information Dissemination in Pervasive RecSys's]{On Gossip-based Information Dissemination in\\Pervasive Recommender Systems}

\author {Tobias Eichinger}
\email{tobias.eichinger@tu-berlin.de}
\orcid{0000-0002-8351-2823}
\affiliation{%
	\institution{Service-centric Networking}
	\institution{Technische Universit\"{a}t Berlin}
	\streetaddress{Straße des 17. Juni 135}
	\city{Berlin}
	\state{Germany}
	\postcode{10623}
}

\author {Felix Beierle}
\email{beierle@tu-berlin.de}
\orcid{0000-0003-2702-9893}
\affiliation{%
	\institution{Service-centric Networking}
	\institution{Technische Universit\"{a}t Berlin}
	\streetaddress{Straße des 17. Juni 135}
	\city{Berlin}
	\state{Germany}
	\postcode{10623}
}

\author{Robin Papke, Lucas Rebscher, Hong Chinh Tran, Magdalena Trzeciak}
\affiliation{%
	\institution{Technische Universit\"{a}t Berlin}
	\streetaddress{Straße des 17. Juni 135}
	\city{Berlin}
	\state{Germany}
	\postcode{10623}
}

\renewcommand{\shortauthors}{Eichinger et al.}

\begin{abstract}
Pervasive computing systems employ distributed and embedded devices in order to raise, communicate, and process data in an anytime-anywhere fashion. Certainly, its most prominent device is the smartphone due to its wide proliferation, growing computation power, and wireless networking capabilities. In this context, we revisit the implementation of digitalized word-of-mouth that suggests exchanging item preferences between smartphones offline and directly in immediate proximity. Collaboratively and decentrally collecting data in this way has two benefits. First, it allows to attach for instance location-sensitive context information in order to enrich collected item preferences. 
Second, model building does not require network connectivity. Despite the benefits, the approach naturally raises data privacy and data scarcity issues. In order to address both, we propose Propagate and Filter, a method that translates the traditional approach of finding similar peers and exchanging item preferences among each other from the field of decentralized to that of pervasive recommender systems. Additionally, we present preliminary results on a prototype mobile application that implements the proposed device-to-device information exchange. Average ad-hoc connection delays of 25.9 seconds and reliable connection success rates within 6 meters underpin the approach's technical feasibility.
\end{abstract}



\maketitle

\section{Introduction}\label{section:introduction}
Nowadays, the possibility to collect, store, and process very large amounts of data in combination with powerful data transformation and analysis techniques have raised privacy concerns such as when in early 2018, Cambridge Analytica had been granted access to millions of Facebook user profiles for political campaigning without users being aware of it. Statutory counter measures to curb data misuse have in particular been undertaken by the European Union in May 2018 by adopting the General Data Protection Regulation 
(GDPR). The GDPR imposes strict rules on data processing, ownership, information obligation (including raising consent), transparency, and collection of user-related data. Since data protection laws penalize misuse of personal data reactively, they do not actively prohibit misuse on a technical level. 
In the literature, three major recommender system architectures have been proposed to address privacy issues arising from collecting large authoritative data pools.

Federated learning produces personalized recommendation models by communicating model building between a central server and mutually disconnected peers holding personal information \cite{McMahan2017,Chen2018}. Thus, federated learning enacts model building centrally on data that is distributed across personal data owners.

In contrast, decentralized recommender systems feature direct interactions between distributed peers without a central server. They are commonly built on top of file-sharing peer-to-peer networks \cite{Baraglia2013,Magureanu2012,Ruffo2009,Ziegler2005} that use gossip mechanisms \cite{Jelasity2007} in order to establish a logic overlay network for fast network search and network resilience in view of peers joining or churning the network. 
In short, dissimilar peers are dropped from and similar peers are added to views (lists of visible peers in the network) iteratively. In so doing, the network establishes homogenous interest groups, which share recommendations explicitly among each other.

Pervasive (or ubiquitous) recommender systems \cite{Polatidis2015,Mettouris2014} are systems that often revolve around location-aware recommendations of for instance items in a nearby shop, restaurants, or events in proximity. 
The location-based approach naturally circumvents transmitting personal data to a central remote authority for recommendation by instead requesting closeby profile as well as context information. Pervasive recommender systems are naturally confronted with limited profile data, for usually only a small subpopulation is available in proximity. Fortunately, there are indications that integrating context data into the recommendation process allows to outweigh profile data scarcity \cite{Adomavicius2005}. 

Among these three approaches we believe that pervasive recommender systems yield the highest potential for data privacy for two reasons. First, the availability of context data as well as mobile compute power on smartphones increases rapidly rendering more and more complex recommendation algorithms feasible on smartphones holding personal data. Second, the model building process happens on-device and does not require connectivity to the network. Consequently, peers become invisible in the network when they do not interact with other peers thus adding privacy on the level of model building. In spite of their potential for recommendation, we see two main burdens. First, pervasive recommender systems are susceptible to data privacy issues since the recommendation mechanism usually builds on the exchange of raw profile data with nearby peers. Second, local scarcity of profile data renders item recommendations on location-independent items taxing. We believe that introducing gossip-based mechanisms and data sampling strategies to the field of pervasive recommender systems can alleviate both issues.

We present the following preliminary results:
\begin{itemize}
	\item The design of \textit{Propagate and Filter}, a gossip-based method 
	that addresses the problem of data privacy and data scarcity in pervasive recommender systems. 
	\item  An implementation of the propagation part in the form of an Android mobile application utilizing Google's Nearby Connections API and its evaluation.
\end{itemize}

\section{Related Work}\label{section:relatedwork}
In the present work, we propose a method for disseminating recommendations epidemically in combination with an on-device filtering process for which we proposed a mobile software architecture in \cite{Beierle2019UIC}. The most similar work is that of Barbosa et al. \cite{Barbosa2018} that propose device-to-device raw profile exchanges in an opportunistic networking scenario. Yet, it differs in that they neither address scalability nor privacy issues. Other related work subsumes:

\textit{Information Dissemination}: 
	Technical works on device-to-device information exchange in proximity are key enablers for collaboratively built recommender systems on mobile devices.
	The \textit{Haggle} API \cite{Nordstrom2009} implements data-centric 
	message forwarding using Bluetooth, Ethernet, and WiFi\footnote{\url{http://user.it.uu.se/~erikn/papers/haggle-arch.pdf}}. 
	In \cite{PietilainenMobiCliqueMiddlewareMobile2009}, the
	authors implement a Bluetooth-based middleware to create opportunistic networks between passing users. 
	On top of such enablers, the authors in  \cite{BoutetC3PONetworkApplication2015a} leverage real-world social phenomena such as large crowds at sports events for event-based mobile communication. 

\textit{Information Filtering}: Many approaches to collaboratively and decentrally build recommender systems on mobile devices in pervasive computing environments disseminate information epidemically and generate recommendations on top of user similarity. User similarities have been found to be reliably estimated by contextual information in the form of proximity at a music festival \cite{deSpindler2007} or spatio-temporal information in the tourism domain \cite{Gavalas2014}. Estimating user similarity on rating vectors has been performed on subcommunities \cite{Delprete2010} or affinity networks \cite{Schifanella2008,Veradelcampo2012}.

\section{Design}\label{section:design}

At the proposed information dissemination method's core is the social movement of people carrying mobile devices. On top of that core movement, we conceive a background data exchange between devices whenever they are geographically close to each other followed by an on-device customizable filtering process. We call this method \textit{Propagate and Filter}.

\subsection{Similarity Data, Peer Preference List, and Neighborhood Preference List}\label{section:design_data}

\begin{figure}[htbp]
	\includegraphics[width=\linewidth,trim={6cm 0 5.5cm 0},clip=true]{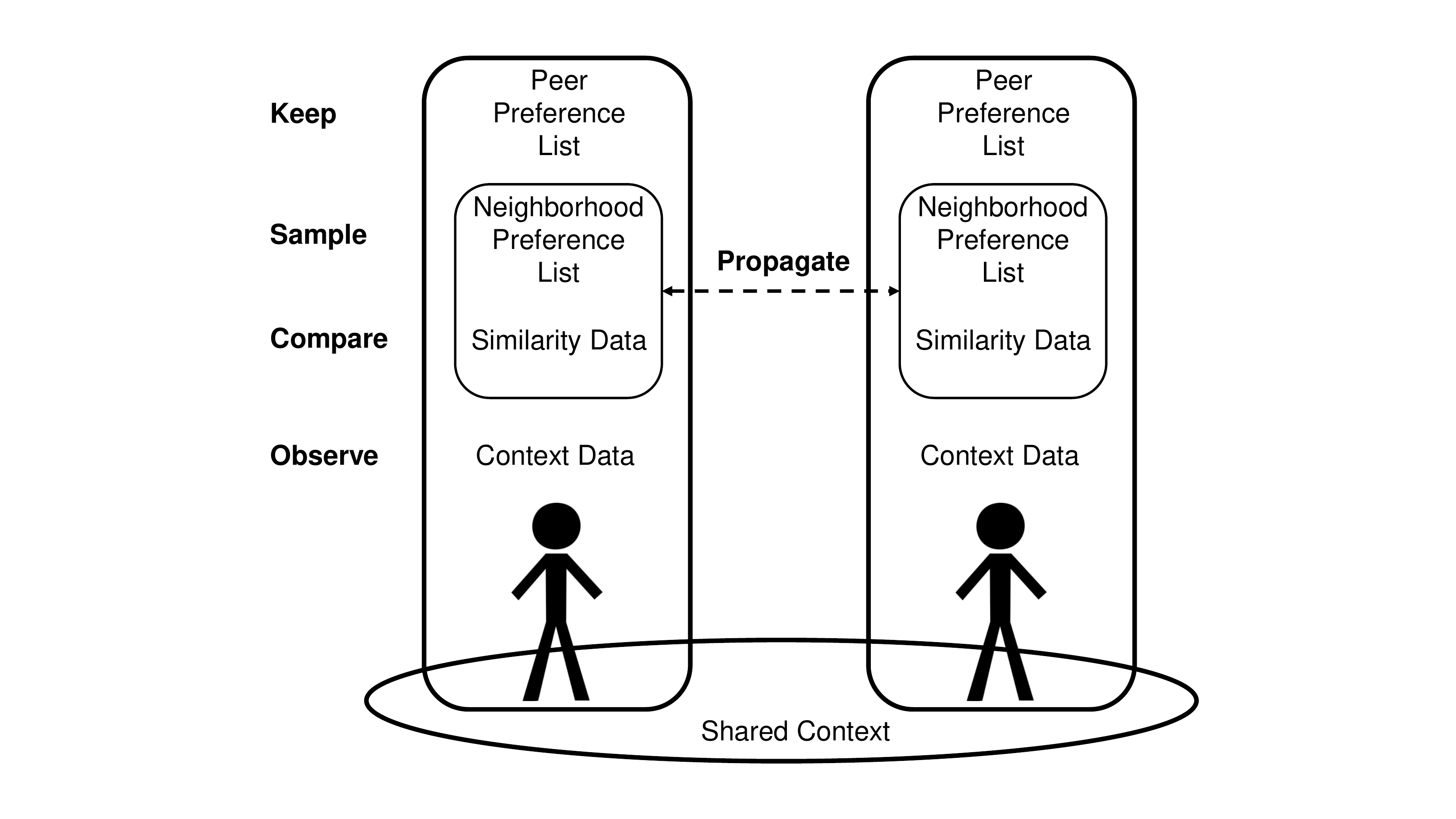}
	\caption{Four types of data proposed by Propagate and Filter and their respective use at propagation. Recommendations are calculated on the mobile device upon all four data pools.}
	\label{fig:design_data}
\end{figure}

Each device carries four types of data (see Figure \ref{fig:design_data}), where the entire stack of data\footnote{The integration of retrievable content information on item references received from nearby peers is omitted here since it is independent of pervasive recommender systems.} is used for the derivation of local personalized recommendations.

\begin{itemize}

	\item \textit{Peer Preference List}: A list of items rated by the peer\footnote{Commonly referred to as \textit{item vector}, yet omitted due to a conceptual overlap with neighborhood preference list.}. It can contain binary or scalar ratings. In our prototype implementation, the peer can rate movies, where each movie is identified with a unique identifier provided by the publicly available Internet Movie Database\footnote{https://www.imdb.com/} (IMDb). Peer preference lists are kept on the device.

	\item \textit{Neighborhood Preference List}: Every peer mixes\footnote{We envision a bootstrapping approach for mixing \cite{Breiman1996,Efron1979}.} previously collected neighborhood preference lists received from the $k$ most similar peers into a single list of item ratings. It is thus an aggregated preference list of an unknown subset of peers. Neighborhood preference lists get propagated to other peers. Note that every peer controls the amount of his/her own peer preference list that gets propagated to nearby peers.

	\item \textit{Similarity Data}: Any kind of data that can be used for peer similarity comparison. Similarity data gets propagated to nearby peers and therefore has to be privacy-preserving. Privacy-preserving similarity comparison can among others be performed on item vectors \cite{BeierleYouWhatSimilarity2018} as well as texting data \cite{Eichinger2019}. 

	\item \textit{Context Data}: Data that characterizes the encounter such as location, time, weather, or peer activity (running, eating, commuting) that can be \textit{sensed} (for example via sensors) or \textit{retrieved} (for example from the web) \cite{Beierle2018MobiSPC,Yurur2014}.
\end{itemize}

\subsection{Propagate}\label{section:design_propagate}
When two or more peers are geographically close to each other, their smartphones establish pairwise fast and secure connections and exchange their neighborhood preference lists and similarity data (propagation). At propagation time the received data is enriched with context data such as time or location characterizing the encounter. 

\subsection{Filter}\label{section:filter}
Data collection in the propagation step includes unfiltered data from every encounter. It is necessary to filter by similarity in order to arrive at relevant information. Upon receiving data from another peer, the filtering process starts on the device. Three steps happen:

\begin{enumerate}
	\item \label{section:simcomp}\textit{Similarity Comparison}: Similarity data is used in order to compare peer similarity between sender and receiver. 

	\item \label{section:resample} \textit{Sample Neighborhood Preference List}: If the peer similarity is above the $k$-th highest, resample the neighborhood preference list on the basis of the peer's preference list and the neighborhood preference lists by the $k$ most similar peers.

	\item \label{section:recommend} \textit{Update Personal Recommendations}: Run a recommendation algorithm\footnote{Propagate and Filter is independent of any specific recommendation algorithm. The filtering techniques described in \cite{Adomavicius2011} are viable approaches leveraging contextual data.} on the locally available data (see Figure \ref{fig:design_data}) in order to derive new recommendations or update ratings of previously generated recommendations.
\end{enumerate}

\subsection{Privacy by Disconnection}
Propagate and Filter's propagation step establishes wireless connections between smartphones in proximity in an ad-hoc fashion, exchanges similarity data and neighborhood preference lists, and thereafter terminates the connection. Consequently, the network topology is essentially disconnected and no information on inter-peer relationships -- such as is the case with Peer-to-Peer network overlays in social networks \cite{Ktari2011}
, recommender systems \cite{Baraglia2013}, or vehicular networks \cite{Heep2013} -- are exploitable. The entire peers' local databases form a geographically distributed and essentially disconnected database, where every peer holds only a very limited portion of data. 
Data query and data search are thus not possible at will. Access to other peers' data is limited to the time of contact and amount of data individually made available by nearby peers. We call this property \textit{Privacy by Disconnection}. It is Propagate and Filter's contribution to data privacy.

\subsection{Enhance Locally Available Profile Data}
Gossip protocols require a connected peer-to-peer network in order to converge similar peers toward each other, where network connectivity is retained by peer sampling \cite{Jelasity2006}. In traditional decentralized recommender systems, peer sampling that requires network connectivity is applicable since neither items nor peers are \emph{spatial}, that is both can be moved in the network at will. In the Propagate and Filter scenario, peers \emph{are} spatial and cannot be moved in the network at will, though items can. Therefore, Propagate and Filter proposes to converge the recommendation list of latent interest communities in the following way:
Peer sampling does not have to be administered by any protocol since it is performed by the global movement of self-organized agents carrying smartphones. Propagate and Filter unfolds its dissemination potential at locations of high degrees of peer mobility such as urban areas \cite{ChancayGarcia2018}. When a peer receives data from a similar peer, he/she resamples his/her neighborhood preference list and else does nothing. As a consequence, Propagate and Filter creates a constant flow of item recommendations that flows between similar peers and dries out between dissimilar peers. This property allows to relay recommendations between peers that have never been geographically close to each other and avoids having to transmit any information on dissimilar peers. In that sense Propagate and Filter addresses the profile data scarcity problem prone to pervasive recommender systems.

\section{Implementation}\label{section:implementation}

We present an Android mobile application that implements a prototype of the propagation step described in Section \ref{section:design_propagate}. We restrict to the transmission of peer preference lists for reasons of simplicity. The application is 
available in the Google Play Store\footnote{\url{https://play.google.com/store/apps/details?id=de.tub.affinity.android}}.

\subsection{Prepare a Peer Preference List}
The prototype allows to search and rate movies locally which are registered and uniquely identified in the IMDb. Movie ratings are scalar, ranging from 1 to 5 stars, and stored in the format (userID, movieID, scalarRating). A list of movie ratings implements the peer preference list introduced in Section \ref{section:design_data}. Once ratings have been specified by the user, he/she can activate sharing. 

\subsection{Propagate via Google Connections API}
The application's active sharing mode applies \textit{advertise} -- the device broadcasts its existence to other devices in proximity -- and \textit{discover} -- the device listens on other device's broadcasting. The sharing process is handled by  the Google Nearby Connections API\footnote{\url{https://developers.google.com/nearby/connections/overview}}  and entirely happens in the background. The Google Nearby Connections API connects devices using one of the three wireless technologies Bluetooth, Bluetooth Low Energy (BLE), or WiFi, automatically selecting the most efficient one in each scenario.

When two smartphones establish a connection, ratings are exchanged immediately and stored locally. Note that ratings can be exchanged without a connection to the internet as it is commonly the case in underground trains. As soon as the internet connection is re-established, detailed movie information such as the movie genre, cast or, trailer can be queried via movieIDs.

\section{Evaluation}\label{section:evaluation}

We evaluate the prototype implementation in view of its potential to facilitate the proposed pervasive recommender system sketched in Section \ref{section:design} in particular in urban areas. We conducted six distinct experiments, where every experiment was conducted with Nexus 5 smartphones that support Bluetooth up to version 4 (including BLE) and Wifi direct. We re-ran experiments 10 times by default, unless stated otherwise, with distinct experiment setups in order to have empirical evidence of the results' soundness.

\begin{enumerate}
	\item \label{subsection:largeamounts} \textit{Share Large Amounts of Ratings}: We share a bulk of 1000 ratings, accounting to roughly $100$ kB uncompressed. As soon as the connection is established, all 1000 ratings get transmitted instantly and reliably without any data loss.

	\item \label{subsection:multipledevices} \textit{Share Data Between Multiple Devices}: We share ratings between four smartphones simultaneously holding mutually disjoint ratings. 
	Ratings get transmitted correctly and losslessly.
	
	\item \label{subsection:publictransportation} \textit{Share Data in Public Transportation}: We successfully share ratings between three devices in the bus and the underground in the urban city of Berlin being exposed to many WiFi and Bluetooth disturb signals. The loss of internet connectivity does not impair the data transfer. As soon as internet connectivity is re-established, the shared placeholder recommendations get filled with 
 movie metadata fetched from the Open Movie Database\footnote{\url{http://www.omdbapi.com/}} (OMDb) API.
 
	\item \label{subsection:range} \textit{Effective Transmission Range}: Recall that the Nearby Connections API comprises Bluetooth, 
	BLE, and Wifi direct under its hood. All three allegedly provide an effective transmission radius of $10$ meters for class 2 devices such as smartphones. 
We tested transmission ranges between $3$ and $12$ meters, both outdoors and indoors, and with and without obstacles. The results are shown in Table \ref{tab:range}. 
	We conclude that the effective radius of ratings propagation is between $3$ and $6$ meters.

	\begin{table}[t]
	\centering
	\caption{Connection success rates subject to transmission range}
	\label{tab:range}
	\begin{tabular}{@{}rrr@{}}
		{} & \multicolumn{2}{c}{\textbf{success rate}}  \\
		\cmidrule(l){2-3}
		\textbf{distance} &  \textbf{without obstacles} & \textbf{with obstacles} \\
		\midrule
		3 m & 100\% & 100\% \\
		6 m & 80\% & 70\% \\
		10 m & 20\% & 0\% \\
		12 m & 0\% & 0\% \\
		\bottomrule
	\end{tabular}
	\\[6pt]
	\end{table}
	\item  \label{subsection:delay} \textit{Average Initial Connection Delay}: We measure the initial connection delay for two devices, that is the elapsed time before a connection gets established. We place the devices at a distance of $1$ meter in order to guarantee connectivity, where the connections were made with distinct app sessions (partner device's id not in cache). The average connection delay is $25.9$ seconds, with a minimum and maximum connection delay of $11$ and $41$ seconds respectively.

	\item \label{subsection:battery} \textit{Battery Drainage}: Since the advertising and discovery step (pre-connection) has to be performed continuously, we would like to know whether or not advertising, discovering, and sharing information in the background for extended periods of time is feasible or not. Since we cannot experiment in a real-life scenario with at times no devices to connect to, and then multiple devices to connect to, we only measure the application's pre-connection battery drainage, which therefore presents a lower bound for battery drainage. We use two devices reset to factory settings. We track the battery levels for three distinct scenarios (a) application running in the background with sharing on, (b) with sharing off, and (c) factory settings, where in all three cases the displays were off. The results are shown in Table \ref{tab:battery}.

\end{enumerate}

\begin{table}[t]
	\centering
	\caption{Average battery drain of pre-connection (advertising and discovery)}
	\label{tab:battery}
	\begin{tabular}{lrrr}
		\midrule
		\textbf{scenario} &  \textit{Battery Drain} & \textit{Battery Logger} & Average \\
		\midrule
		sharing on  & 5.94 \%/h & 5.59 \%/h & 5.77 \%/h \\
		sharing off & 0.34 \%/h & 0.66 \%/h & 0.50 \%/h \\
		factory settings      & 0.51 \%/h & 0.49 \%/h & 0.50 \%/h \\
		sharing only    & 5.6 \%/h  & 4.93 \%/h & 5.27 \%/h \\
	\end{tabular}
	\\[6pt]
\end{table}

\subsection{Conclusion}
The experimental findings indicate that the Propagate and Filter's propagation step works reliably with larger amounts of ratings, in multi-device scenarios, and in areas without internet connectivity such as underground trains. Yet, the implementation has certain limitations. First of all, information can only be disseminated reliably within a radius of $6$ meters, which on the one hand strengthens privacy, and on the other limits the number of potential peers to share information with. Furthermore, the average initial connection delay of $25.9$ seconds is considerable and does not allow to share information between for instance passing pedestrians, yet includes scenarios such as waiting at the traffic lights, sitting next to each other in a caf{\'{e}} or restaurant, or taking public transportation. Last but not least, battery drainage is relatively high at at least $5$\% per hour, which thus prohibits continuous advertising and discovery. Apart from ever improving transmission standards and hardware, we believe that it is possible to limit activity and inactivity of the sharing process efficiently on the logical level leveraging sensor information on smartphones.

\section{Future Work}\label{section:futurework}
Future work includes the launch of the mobile application for the collection of usage data. Recall that we left the sampling process and the recommendation algorithm as placeholders (see Section \ref{section:filter} (\ref{section:resample}) and (\ref{section:recommend}) respectively). Once usage data is collected it will be possible to test distinct combinations of sampling and recommendation strategies. Furthermore, the evaluation of feedback on the application including the user interface and user experience is due.

\newpage

\bibliographystyle{ACM-Reference-Format}

\newpage
\includepdf[pages=-,pagecommand={},width=\textwidth]{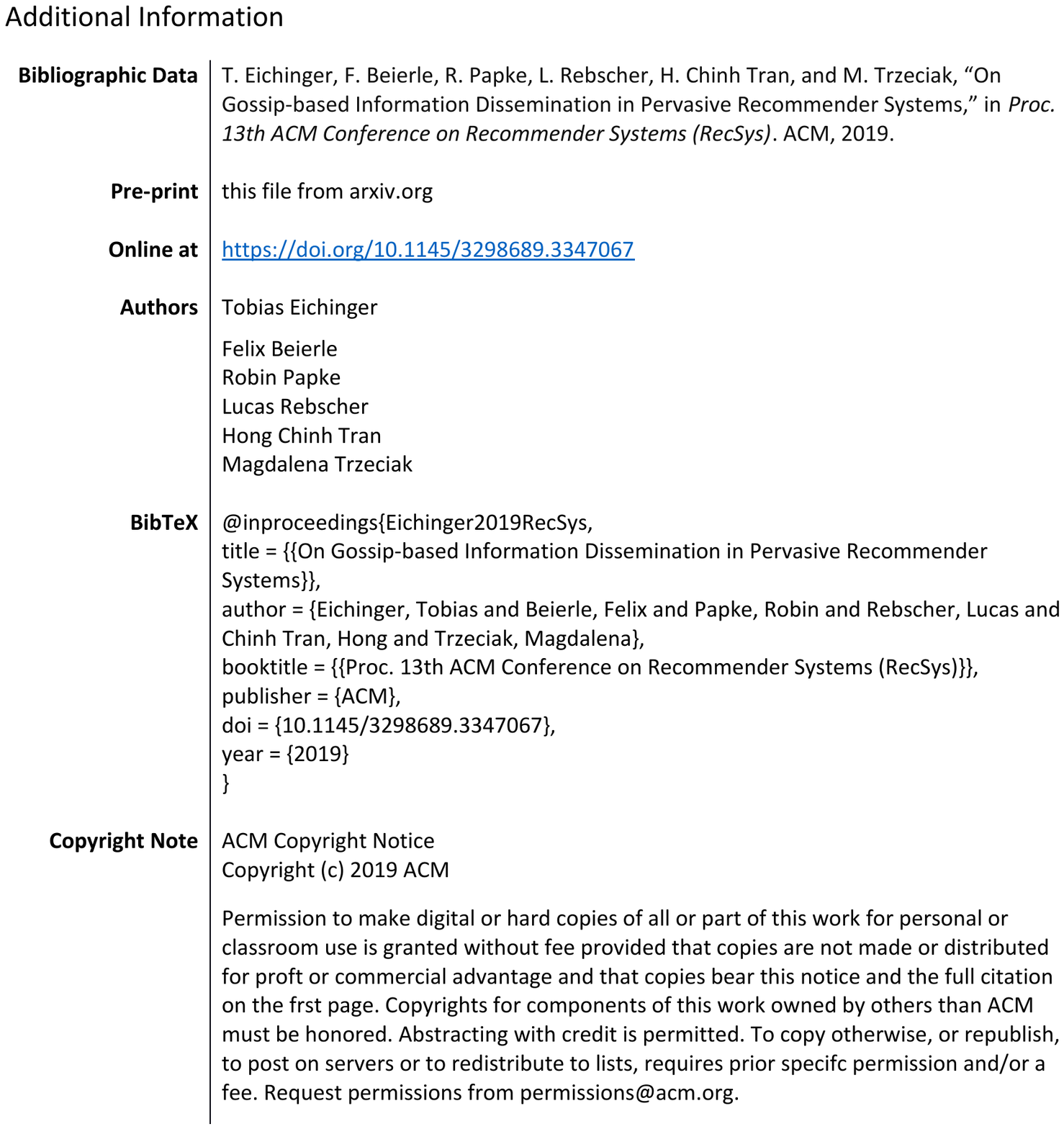}

\end{document}